\documentclass[prb,twocolumn,aps,showpacs,fixfloats]{revtex4}
\usepackage{graphicx}
\usepackage{subfigure}
\usepackage{float}
\usepackage{bm}
\usepackage{amsmath,amssymb}
\DeclareMathOperator{\sign}{sign}
\usepackage{latexsym}
\usepackage{color}
\usepackage{enumerate}
\usepackage{pdfpages}
\usepackage{tikz}
\usepackage{hyperref}

\begin{document}
\newcommand{\s}{\scriptscriptstyle}
\newcommand{\uu}{\uparrow \uparrow}
\newcommand{\ud}{\uparrow \downarrow}
\newcommand{\du}{\downarrow \uparrow}
\newcommand{\dd}{\downarrow \downarrow}
\newcommand{\ket}[1] { \left|{#1}\right> }
\newcommand{\bra}[1] { \left<{#1}\right| }
\newcommand{\bracket}[2] {\left< \left. {#1} \right| {#2} \right>}
\newcommand{\vc}[1] {\ensuremath {\bm {#1}}}
\newcommand{\tr}{\text{Tr}}
\newcommand{\Trans}{\ensuremath \Upsilon}
\newcommand{\Refl}{\ensuremath \mathcal{R}}

\title{Resonant reflection of interacting electrons from an impurity in a quantum wire: interplay of Zeeman and spin-orbit effects}

\author
{Rajesh K. Malla   and M. E. Raikh}

\affiliation{ Department of Physics and
Astronomy, University of Utah, Salt Lake City, UT 84112}
\begin{abstract}
A single-channel quantum wire with two well-separated Zeeman subbands and in
the presence of a weak spin-orbit coupling is considered. An impurity level which is split off the upper subband is degenerate with the continuum of the lower subband. We show
that, when the Fermi level lies in the vicinity of the impurity level, the transport
is completely blocked. This is the manifestation of the effect of resonant reflection
and can be viewed as resonant tunneling between left-moving and right-moving electrons via the impurity level. We incorporate electron-electron interactions and study their
effect on the shape of the resonant-reflection profile. This profile becomes a two-peak structure, where one peak is caused by resonant reflection itself, while
the origin of the other peak is reflection from the Friedel oscillations of the
electron density surrounding the impurity.

\end{abstract}

\maketitle

\section{Introduction}
Electron states in a ballistic wire in the presence of spin-orbit coupling
became the subject of intensive theoretical, see e.g. Refs. \onlinecite{Moroz1999, Pershin2004, Sanchez2006, Sanchez2008, Inhomogeneous, Sherman2017}, and experimental\cite{Goldhaber,InSb2016,InSb2017,InAs2018} studies almost three decades ago. Initial motivation for these studies was  the
proposal of a spin transistor by Das and Datta.\cite{DasDatta}
The motivation for the later studies
was the proposal\cite{Oreg,DasSarma} that, in the proximity to a superconductor, the interplay of spin-orbit coupling and Zeeman splitting
can lead to the formation of zero-energy bound states at the wire ends.
Yet another motivation for the research on the combined action of Zeeman and spin-orbit fields comes from the recent experiments on cold gases.\cite{Spielman2011}

Nontriviality of the interplay of spin-orbit coupling and Zeeman splitting manifests
itself already in the ballistic transport through the wire. It was predicted\cite{Moroz1999,Pershin2004} and confirmed experimentally\cite{Goldhaber}
that, as a result of this interplay, the dependence of the conductance on the
Fermi level can become non-monotonic. Such a ``spin gap" develops when the spin-orbit minimum in energy spectrum of a free electron is comparable to the Zeeman splitting.
Another nontrivial consequence of the interplay shows up when the spin-orbit coupling is inhomogeneous\cite{Sanchez2006,Sanchez2008,Inhomogeneous,Sherman2017}.
Namely, a step-like inhomogeneity can lead to a full reflection of the incident electron.

\begin{figure}[h!]
\includegraphics[scale=0.2]{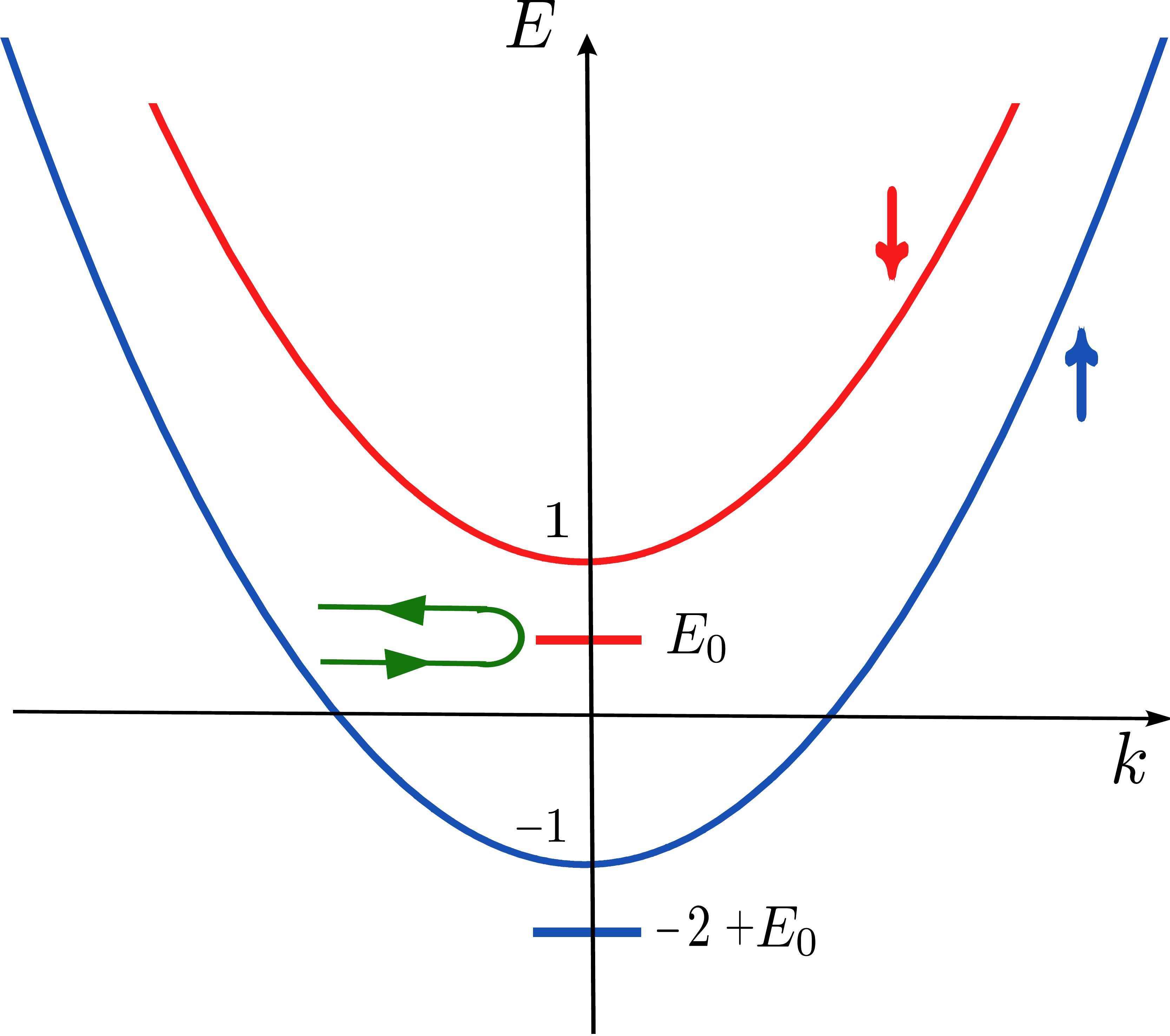}
\caption{(Color online) Schematic illustration of the resonant reflection. An attractive impurity creates bound states
under the bottoms of $\downarrow$ (red) and $\uparrow$ (blue) sub-bands.
The binding energy, measured in the units of $\Delta$, is $1-E_{\s 0}$
Weak spin-orbit coupling mixes $\downarrow$ and $\uparrow$ wave functions. As a result, an incident $\uparrow$ electron undergoes a resonant scattering, illustrated by a green line.
The result of the scattering is almost full reflection rather than conventional resonant transmission.   }
\label{firstfigure}
\end{figure}

The underlying physics of the full reflection is the same as the physics of the resonant reflection in the two-subband wire first studied in Refs. \onlinecite{Levinson1993}, \onlinecite{Stone}. It does not require either Zeeman field or spin-orbit coupling.
An attractive impurity in a two-subband wire splits off an energy level from the
bottom of both subbands. If the Fermi level, lying in the lower subband,
coincides with the level split from the upper subband, see Fig. \ref{firstfigure}, the transport involves multiple virtual visits to this level. As it was first shown in Ref. \onlinecite{Levinson1993}, the outcome of these visits
is a reflection rather than resonant transmission as one would naively expect.
In a single-channel wire the role of the size-quantization subbands is played by
the spin subbands, while the visits to the split-off level are enabled by the spin-orbit coupling.

The goal of the present paper is to study the effect of electron-electron interactions on the resonant reflection. For a single-channel interacting wire it is accepted
that any weak potential impurity blocks completely the zero-temperature transport through the wire.
The theories\cite{FisherReview} which capture this phenomenon are
Luttinger-liquid description and backscattering by the Friedel oscillations in electron gas imposed by an impurity. In the latter case, the role of interactions
is simply a conversion of the oscillations of electron density into the oscillations of the potential. As it was first pointed out in Refs. \onlinecite{Yue1993, Yue1994}
(see also later papers Refs. \onlinecite{Maslov, Nagaev}), the period of the Friedel
oscillations matches the Bragg condition for electron at the Fermi level. Thus, the
electron is scattered by a compound object consisting of the impurity itself and the oscillating potential, which it creates.

The theory of Refs. \onlinecite{Yue1993, Yue1994} was later generalized to the case
of a pair of impurities.\cite{Nazarov2003,Gornyi2003} Specifics of the pair is that
electron can bounce between the constituting impurities for a long time. As a result
of this bouncing, a quasi-local level degenerate with the continuum is formed. For incident electron with energy in resonance with this quasi-local level  the transmission coefficient is close to $1$. Physically, the results of
Refs.  \onlinecite{Nazarov2003, Gornyi2003} can be interpreted as follows.
When the incident electron is resonantly transmitted, the Friedel oscillations do
not form, so that the interactions suppress the transmission only when the Fermi
level is spaced away from the resonant level.

Contrary to the resonant transmission, in the case of the resonant
reflection the Friedel oscillations are the strongest when the Fermi
level lies close to the impurity level. Thus, the modification of the
resonant reflection profile due to interactions is also strong. This
demands a more detailed treatment of partial reflection of electron on the
way to the impurity than the renormalization-group scheme adopted in
Refs. \onlinecite{Yue1993, Yue1994, Maslov, Nagaev, Nazarov2003, Gornyi2003}.
Our most spectacular finding is that, for certain phases accumulated by the electron on the way to the impurity, the resonant reflection from the bare impurity can turn
into the resonant transmission.

\section{Resonant reflection}

In the presence of the Zeeman field and spin-orbit coupling, the
Hamiltonian of a wire has the form
\begin{equation}
\label{Hamiltonian}
\hat{H}=\begin{pmatrix}
-\frac{\hbar^2k_x^2}{2m}-\Delta && i\gamma k_x \\ \\
-i\gamma k_x && -\frac{\hbar^2k_x^2}{2m}+\Delta
\end{pmatrix},
\end{equation}
where $m$ is the electron mass, $2\Delta$ is the Zeeman splitting, and $\gamma$
is the spin-orbit coupling strength.

We assume that the impurity potential is short-ranged, $V(x)=V_0\delta(x)$.
The system of coupled equations for $\uparrow$ and $\downarrow$ components of the spinor reads
\begin{eqnarray}
\label{System1}
-\frac{\hbar^2}{2m}\frac{\partial^2\psi_1}{\partial x^2}+V_0 \delta (x)\psi_1 - (\varepsilon +\Delta)\psi_1=\gamma\frac{\partial \psi_2}{\partial x},\nonumber \\
-\frac{\hbar^2}{2m}\frac{\partial^2\psi_2}{\partial x^2}+V_0 \delta (x)\psi_2 - (\varepsilon -\Delta)\psi_2=-\gamma\frac{\partial \psi_1}{\partial x}.
\end{eqnarray}

Since the energy of the incident $\uparrow$ electron in resonance with impurity
level of $\downarrow$ electron is close to $\Delta$, see Fig. \ref{firstfigure},
it is convenient
to introduce the following dimensionless variables
\begin{eqnarray}
\label{Variablechange}
 z=\frac{x}{x_0}, \hspace{1cm} E=\frac{\varepsilon}{\Delta}, \nonumber  \\
 \alpha=\left(\frac{2m x_0}{\hbar^2}\right)\gamma, ~U_0=\left(\frac{2m x_0}{\hbar^2}\right)V_0,
\end{eqnarray}
where the characteristic length,
\begin{equation}
\label{x0}
x_0=\left(\frac{\hbar^2}{2m\Delta}\right)^{1/2},
\end{equation}
is the de Broglie wave length of the electron with energy $\varepsilon =\Delta$.
In  the dimensionless  variables the system Eq.~(\ref{System1}) takes the form
\begin{eqnarray}
\label{System2}
-\frac{\partial^2\psi_1}{\partial z^2}+U_0 \delta(z)\psi_1 - (E +1)\psi_1=\alpha\frac{\partial \psi_2}{\partial z},\nonumber\\
-\frac{\partial^2\psi_2}{\partial z^2}+U_0 \delta (z)\psi_2 - (E -1)\psi_2=-\alpha\frac{\partial \psi_1}{\partial z}.
\end{eqnarray}
Without impurity, the solutions of the system Eq. (\ref {System2}) in the domain $-1<E<1$ correspond to propagation of $\uparrow$ spin component and  the decay of $\downarrow$ spin component, see Fig. \ref{firstfigure}. Due to spin-orbit coupling,
both components of corresponding spinors are nonzero,
\begin{equation}
\label{wavefunctions}
\begin{pmatrix}
\psi_1 \\ \\ \psi_2
\end{pmatrix}=\begin{pmatrix}
1 \\ \\ iC
\end{pmatrix}e^{iqz},~\begin{pmatrix}
\psi_1 \\ \\ \psi_2
\end{pmatrix}=\begin{pmatrix}
D \\ \\ 1
\end{pmatrix}e^{-\kappa z},
\end{equation}
where the wave vector, $q$, the decay constant, $\kappa$, and the components, $C$ and $D$, of the spinors  are given by
\begin{eqnarray}
\label{q}
q(E)&=&(1+E)^{1/2},~ \kappa(E)=(1-E)^{1/2}, \nonumber \\
 C&=&\frac{1}{2}\alpha q,~~~~D=\frac{1}{2}\alpha\kappa.
\end{eqnarray}
Coefficients $C$ and $D$ describe the admixture of the opposite spin projection due to spin-orbit coupling.

In the presence of impurity, the general solution at $z<0$ has a form
\begin{equation}
\label{Negativez}
\begin{pmatrix}
\psi_1 \\ \\ \psi_2
\end{pmatrix}=\begin{pmatrix}
1 \\ \\ iC
\end{pmatrix}e^{iqz}+r_1 \begin{pmatrix}
1 \\ \\ -iC
\end{pmatrix}e^{-iqz}+r_2\begin{pmatrix}
D \\ \\ 1
\end{pmatrix}e^{\kappa z},
\end{equation}
which is the combination of the solutions Eq. (\ref {wavefunctions}).
First two terms describe the incident and the reflected $\uparrow$ waves, while the third term describes the solution corresponding to $\downarrow$, which decays at $z\rightarrow -\infty$.

The corresponding solution for $z>0$ reads
\begin{equation}
\label{Positivez}
\begin{pmatrix}
\psi_1 \\ \\ \psi_2
\end{pmatrix}=t_1\begin{pmatrix}
1 \\ \\ iC
\end{pmatrix}e^{iqz}+t_2\begin{pmatrix}
-D \\ \\ 1
\end{pmatrix}e^{-\kappa z}.
\end{equation}
The first term describes the transmitted $\uparrow$ wave, while the second term describes the decay of $\downarrow$ component.

Although the parameters $C$ and $D$ are proportional to $\alpha$, and thus are small due to the weakness of the spin-orbit coupling, it is these admixtures that are responsible for the resonant reflection. To capture this effect, we follow the standard procedure and calculate the reflection and transmission coefficients from the system of boundary conditions at $z=0$.

Continuity of the wave function Eqs. (\ref{Negativez}) and (\ref{Positivez}) yields two conditions
\begin{eqnarray}
\label{Continuity}
1+r_1+r_2 D=t_1-t_2 D,\nonumber \\
iC(1-r_1)+r_2=iCt_1 +t_2.
\end{eqnarray}
The other two conditions come from the discontinuity of the derivatives, $\frac{\partial\psi_1}{\partial z}$ and $\frac{\partial\psi_2}{\partial z}$,  at $z=0$. Integrating the system Eq. (\ref{System2}) near $z=0$, we get
\begin{eqnarray}
\label{Differentiability}
iq t_1+\kappa t_2 D - \Big[iq(1-r_1)+\kappa r_2 D \Big]=U_0(t_1-t_2 D),\nonumber\\
-q C t_1-\kappa t_2-\Big[-q C-q C r_1+\kappa r_2 \Big] =U_0(i C t_1 +t_2). \hspace{3mm}
\end{eqnarray}
Simplifying the above boundary conditions by introducing $R_2=Dr_2$, $T_2=Dt_2$, and $\lambda=CD$, we get
\begin{eqnarray}
\label{bc11}
R_2+T_2=t_1-r_1-1,\nonumber \\
R_2-T_2=i\lambda(t_1+r_1-1),
\end{eqnarray}
\begin{eqnarray}
\label{bc21}
iq(t_1+r_1)-U_0t_1-iq=R_2\kappa-(\kappa+U_0)T_2,\nonumber \\
(\kappa+U_0)T_2+\kappa R_2=-\lambda\Big[-it_1(iq-U_0)-q(1+r_1) \Big].\hspace{3mm}
\end{eqnarray}
Since we are interested in the reflection and transmission coefficients, $r_1$ and $t_1$, it is convenient to express $R_2$ and $T_2$ from the system Eq. (\ref{bc11}) and substitute them into the system Eq. (\ref{bc21}), which assumes the form

\begin{eqnarray}
t_1+r_1=\frac{\Big[q-\lambda(\kappa+\frac{U_0}{2}) \Big]+i\frac{U_0}{2}}{\Big[q-\lambda(\kappa+\frac{U_0}{2}) \Big] -i\frac{U_0}{2}},\label{t1r1} \\
t_1-r_1=
\frac{\kappa+\frac{U_0}{2}+q\lambda-i\lambda\frac{U_0}{2}}{\kappa+\frac{U_0}{2}+q\lambda+i\lambda\frac{U_0}{2}}.\label{t1-r1}
\end{eqnarray}
We see that the absolute values of $t_1+r_1$ and $t_1-r_1$ are equal to $1$. Then it
is convenient to cast the solution of the system Eq. (\ref{t1r1})  into the form
\begin{equation}
\label{phi}
|r_1|^2=\sin^2(\Phi_- - \Phi_+),~~|t_1|^2=\cos^2(\Phi_- - \Phi_+),
\end{equation}
where
\begin{eqnarray}
\label{phi+-}
\Phi_+ = \frac{1}{2}\tan^{-1} \frac{\frac{U_0}{2}}{q-\lambda(\kappa+\frac{U_0}{2})}, \nonumber \\
 ~\Phi_-=\frac{1}{2}\tan^{-1}
\frac{\frac{\lambda U_0}{2}}{\kappa+\frac{U_0}{2} +q\lambda}.
\end{eqnarray}
Until now the calculation was exact. Weakness of spin-orbit coupling, quantified by the condition $\alpha \ll 1$, was used
in the explicit expressions for $q$ and $\kappa$. We will now use this condition to simplify the phases $\Phi_+$ and $\Phi_{-}$.
First, we note that the dimensionless parameter
\begin{equation}
\label{lambda}
\lambda=CD=\frac{1}{4}\alpha^2(1-E^2)^{1/2}
\end{equation}
is quadratic in spin-orbit coupling strength. This allows one to simplify $\Phi_+$ to $\tan^{-1}\left(\frac{U_0}{2q}\right)$. Then $\Phi_+$ can be identified with the
scattering phase of $\uparrow$ electron from the impurity {\em in the absence of spin-orbit coupling}.

Turning to the phase $\Phi_{-}$, we note that the small parameter $\alpha^2$ in the expression for $\lambda$ allows one to neglect the term $q\lambda$ in the denominator. Then we see that, for attractive impurity, $U_0<0$, this denominator turns to zero at energy $E=E_{\s 0}$ determined by the condition

\begin{equation}
\label{resonant}
\kappa (E_{\s 0})=\frac{|U_0|}{2}.
\end{equation}
This condition expresses the fact that
 {\em in the absence of spin-orbit coupling},
the energy position of the level of $\downarrow$ electron in the potential $U_0\delta(z)$ is $E=E_{\s 0}$,
see Fig. \ref{firstfigure}.

To establish the energy width, $\Gamma$, of the resonance,  we recast the expression for
$\tan\Phi_{-}$ into the form

\begin{equation}
\label{resonance1}
\tan \left[\Phi_{-}(E)\right]=\frac{1}{8}\alpha^2 (1-E^2)^{1/2}|U_0|\Bigg[\frac{(1-E^2)^{1/2}+\frac{|U_0|}{2}}{1-E-\frac{U_0^2}{4}}\Bigg].
\end{equation}
Near the resonance, $E=E_{\s 0}=1-\frac{U_0^2}{4}$, the expression Eq. (\ref{resonance1}) assumes the conventional  Breit-Wigner form
\begin{equation}
\label{resonance2}
\tan \left[\Phi_{-}(E)\right]=\frac{\Gamma}{E_{\s 0} - E},
\end{equation}
where $\Gamma$ is given by
\begin{equation}
\label{Gamma}
\Gamma=\frac{\alpha^2}{16}|U_0|^3.
\end{equation}
With binding energy of $\downarrow$ electron being $ \frac{U_0^2}{4}$, we see that the width, $\Gamma$,
is much smaller than this binding energy, which justifies the expansion
near the resonance.

If the bound state in the potential $U_0\delta(z)$ is shallow,
i.e. $U_0\ll 1$,
we can replace $\tan^{-1}$ in expression for $\Phi_+$ by the argument.
After that, the final expression for the energy-dependent reflection coefficient assumes the form
\begin{eqnarray}
\label{reflection}
|r_1(E)|^2&=&\sin^2\left[\tan^{-1} \left(\frac{\Gamma}{E_{\s 0} - E}\right) -\frac{|U_0|}{2q} \right]\nonumber\\
&=&\frac{\Big[\Gamma - \frac{|U_0|}{2q}(E_{\s 0} -E) \Big]^2}{(E_{\s 0} -E)^2 +\Gamma^2}.
\end{eqnarray}
It follows from Eq. (\ref{reflection}) that $|r_1(E)|^2$ has a characteristic Fano shape\cite{Fano}. Near the resonance, $E=E_{\s 0}$, it is a Lorentzian with the width, $\Gamma$. As the energy is swept through $E_{\s 0}$, the reflection coefficient passes through zero
(antiresonace) before returning to its non-resonant value $|r_1|^2=\frac{|U_0|^2}{4q^2}$.


\section{Incorporating the electron-electron interactions}
As it was explained in the Introduction, the effect of interactions
is more pronounced in the case of resonant reflection than in the case of resonant transmission.\cite{Nazarov2003,Gornyi2003}
The reason is that the amplitude of the Friedel oscillations is
proportional to the reflection amplitude\cite{Yue1993,Yue1994}
which, for resonant reflection, is close to $1$. On the other
hand, the Friedel oscillation of electron density creates perturbations
which play the role of the  ``Bragg mirrors" for  incident and
transmitted  electron waves. As a result of Friedel oscillations
being strong, each Bragg mirror is highly ``reflective". This suggests
to incorporate the effect of attenuation, caused by the mirrors, more
accurately than in Refs. \onlinecite{Nazarov2003}, \onlinecite{Gornyi2003}.

The process of electron reflection from a compound object consisting
of three scatterers, two Bragg mirrors and impurity between them,
is illustrated in Fig. \ref{mirror}. The rigorous way to describe this
reflection analytically is by employing the scattering matrices of each
scatterer relating the amplitudes of the incoming and outgoing partial
waves. These matrices are defined as follows

\begin{widetext}

\begin{figure}[h!]
\includegraphics[scale=0.5]{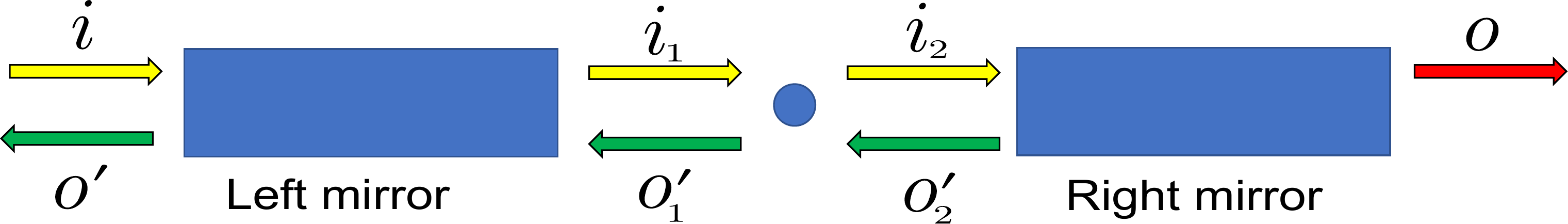}
\caption{(Color online) Schematic illustration of the electron scattering from impurity ``dressed" by Friedel oscillations, which play the role of the Bragg mirrors. The incident electron, $i$, can be reflected by the left mirror, by the impurity, or by the right mirror. }
\label{mirror}
\end{figure}

\begin{equation}
\label{matrices}
\begin{pmatrix}
i_1 \\ \\ o'
\end{pmatrix}=\begin{pmatrix}
t_L && r_L \\ \\ -r_L^* && t_L^*
\end{pmatrix}\begin{pmatrix}
i \\ \\ o'_1
\end{pmatrix},~
\begin{pmatrix}
i_2 \\ \\ o'_1
\end{pmatrix}=\begin{pmatrix}
t_1 && r_1 \\ \\ -r_1^*&& t_1^*
\end{pmatrix}\begin{pmatrix}
i_1 \\ \\ o'_2
\end{pmatrix},~~
\begin{pmatrix}
o \\ \\ o'_2
\end{pmatrix}=\begin{pmatrix}
t_R && r_R \\ \\ -r_R^* && t_R^*
\end{pmatrix}\begin{pmatrix}
i_2 \\ \\ 0
\end{pmatrix}.
\end{equation}
\end{widetext}
The amplitude $r_1$ in Eq. (\ref{matrices}) was found in the
previous section. Two remaining amplitudes, $r_L$ and $r_R$,
will be calculated later. Excluding the intermediate amplitudes,
$i_1$, $i_2$, $o_1'$, $o_2'$ from Eq. (\ref{matrices}), we find
the expression for the net amplitude reflection coefficient of
the compound scatterer
\begin{equation}
\label{reff}
r_{\s eff}=-\frac{o'}{i}=\frac{r_L^*+r_1^*+r_R^*+r_L^*r_1r_R^*}{1+r_1r_R^*+r_Lr_1^*+r_Lr_R^*}.
\end{equation}
To analyze this expression, we express the power reflection coefficient,
$|r_{\s eff}|^2$, via the magnitudes of the reflection coefficients
$r_1$, $r_L$, and $r_R$ and obtain

\begin{equation}
\label{reffabs}
|r_{\s eff}|^2=1-|t_{\s eff}|^2=1-\frac{\left(1-|r_{\textit{\tiny{Bragg}}}|^2\right)^2 \left(1-|r_1|^2\right)}{\Big(1+|r_{\textit{\tiny{Bragg}}}|^2+2|r_{\textit{\tiny{Bragg}}}||r_1|\cos \beta\Big)^2}.
\end{equation}
In Eq. (\ref{reffabs}) we took into account that, unlike Refs. \onlinecite{Nazarov2003}, \onlinecite{Gornyi2003},
there is a symmetry between the left and right mirrors, so that the magnitudes, $|r_L|$ and $|r_R|$
are equal to each other and are denoted with $|r_{\textit{\tiny{Bragg}}}|$. The phase, $\beta$, is the combination
of the phase $\Phi_{-}$, defined by Eq. (\ref{phi}), and the phase, $ \Phi_{\textit{\tiny{Bragg}}}   $,  accumulated in the course of the reflection
from the mirror. We will see that this phase is big and depends strongly on the energy.
Thus, we average Eq. (\ref{reffabs}) over
$\beta$ using the identity
\begin{equation}
\label{identity}
\Bigg{\langle} \frac{1}{(a+\cos\beta)^2} \Bigg{\rangle}_{\beta}=\frac{a}{(a^2-1)^{3/2}}.
\end{equation}
The result of this averaging reads
\begin{equation}
\label{reffavg}
\langle |r_{\s eff}|^2\rangle=1-\frac{\left(1-|r_{\textit{\tiny{Bragg}}}|^2\right)^2\left(1+|r_{\textit{\tiny{Bragg}}}|^2\right) \left(1-|r_1|^2\right)}{\Big[\left(1-|r_{\textit{\tiny{Bragg}}}|^2\right)^2+4|r_{\textit{\tiny{Bragg}}}|^2\left(1-|r_1|^2\right)\Big]^{3/2}}.
\end{equation}
It is also instructive to express the effective power transmission coefficient via the partial
transmission coefficients $|t_1|^2$ and $|t_{\textit{\tiny{Bragg}}}|^2$. One obtains

\begin{equation}
\label{teffavg}
\langle |t_{\s eff}|^2\rangle=\frac{|t_{\textit{\tiny{Bragg}}}|^4\left(2-|t_{\textit{\tiny{Bragg}}}|^2\right) |t_1|^2}{\Big[|t_{\textit{\tiny{Bragg}}}|^4+4\left(1-|t_{\textit{\tiny{Bragg}}}|^2\right)|t_1|^2\Big]^{3/2}}.
\end{equation}

\begin{figure}
\includegraphics[scale=0.31]{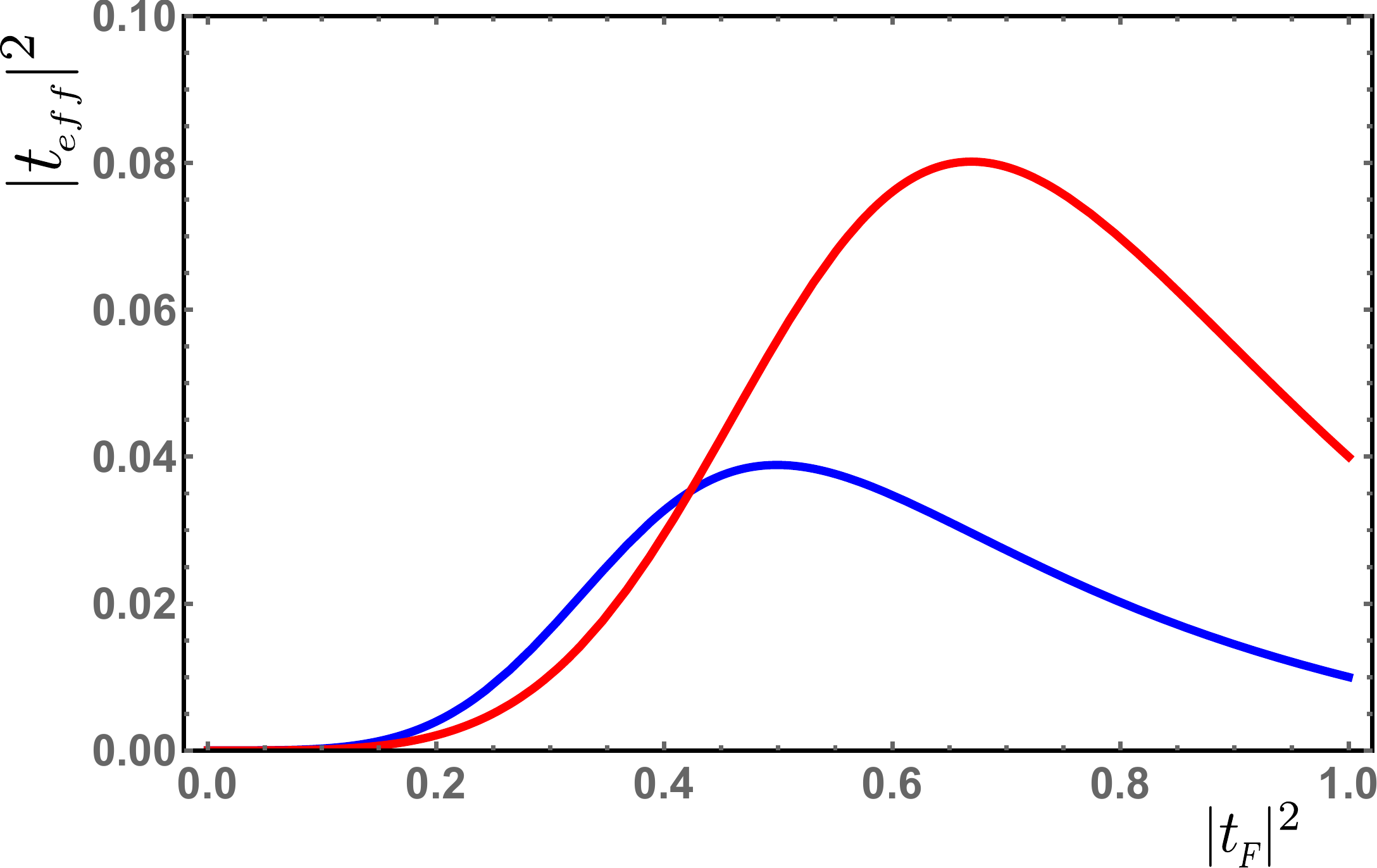}
\caption{(Color online) Effective power transmission coefficient of the impurity dressed by the Friedel oscillations is plotted from Eq. (\ref{teffavg}) versus the transmission of the Bragg mirrors for $|t_1|^2=0.01$ (blue) and $|t_1|^2=0.04$ (red).}
\label{Nonmonotonic}
\end{figure}

Since the transmission, $|t_{\textit{\tiny{Bragg}}}|^2$, is strongly dependent on the position of the Fermi level, $E_{\s F}$, with respect to the resonant energy level, $E_{\s 0}$, the magnitude of $|t_{\textit{\tiny{Bragg}}}|^2$ falls off with increasing $(E_{\s F}-E_{\s 0})$. Then one would expect that $|t_{\s eff}|^2$ grows monotonically with increasing $|t_{\textit{\tiny{Bragg}}}|^2$ and approaches $|t_1|^2$. The reasoning behind this expectation is that
the scattering by the Bragg mirrors becomes inefficient for large $(E_{\s F}-E_{\s 0})$.
Remarkably, the dependence of $|t_{\s eff}|^2$, described by Eq. (\ref{teffavg}), is {\em non-monotonic}.
As illustrated in Fig. \ref{Nonmonotonic}, this dependence has a maximum. For small transmission of the impurity,
$|t_1|^2 \ll 1$, the position of maximum is easy to calculate analytically. It is $|t_{\textit{\tiny{Bragg}}}|^4=8t_1^2$.
Note that the value $|t_{\textit{\tiny{Bragg}}}|^4$ has a meaning of the net transmission of {\em two mirrors}.
Thus, the maximum occurs when the transmissions of the impurity and of the two mirrors are equal within numerical factor. Substituting  $|t_{\textit{\tiny{Bragg}}}|^4=8t_1^2$ into Eq. (\ref{teffavg}), we find the maximal value
of the effective power transmission
\begin{equation}
\left(|t_{\s eff}|^2\right)_{\s max}=\frac{2}{3^{3/2}}|t_1|.
\end{equation}
We see that this value is {\em much bigger} than  $|t_1|^2$.

The origin of the maximum is that the  dominant contribution to the phase-averaged transmission,  $\langle |t_{\s eff}|^2\rangle$, comes from the phases, $\beta$, in  Eq. (\ref{reffabs}) for which the denominator
is close to zero. In other words, while the impurity alone acts as a reflector, adding of the two Bragg mirrors
can lead to the {\em resonant transmission}.

Naturally, the values of $|t_1|^2$ and $|t_{\textit{\tiny{Bragg}}}|^2$ are not independent. It is the reflection from the impurity that controls the magnitude of the Friedel oscillations. To analyze the behavior the effective
transmission with energy, $E$, of the incident electron and with $E_{\s F}$, we need to specify the
analytical form of $|t_{\textit{\tiny{Bragg}}}|^2$. This is done in the next section.
\section{Transmission of the Bragg mirror}
In the presence of electron-electron interactions, propagation of electron through
the mirror is described by the Schr{\"o}dinger equation
\begin{equation}
\label{interaction}
-\frac{\partial^2\psi_1}{\partial z^2}+V_H(z)\psi_1+\hat{V}_{ex} \Big\{\psi_1\Big\}=\left(E+1\right)\psi_1,
\end{equation}
where $V_H(z)$ and $\hat{V}_{ex}$ are the Hartree and the exchange terms, respectively.
When the interaction is short-ranged, one can consider only the Hartree term, since the exchange term
causes only a modification of the interaction constant.\cite{Yue1993} The other consequence
of the interaction being short-ranged is that the Hartree potential is proportional to the
modulation of the electron density created by the Friedel oscillations\cite{Yue1993}, i.e.
it has the from
\begin{equation}
\label{Hartree}
V_H(z)=\frac{\mu(E_{\s F})}{q_{\s F}|z|}\cos(2q_{\s F}|z|),
\end{equation}
where $q_{\s F}$ is the Fermi momentum. The magnitude of the electron-electron interactions
as well as the energy dependence of $|r_1|$, responsible for the Friedel oscillations,  are encoded into
the constant, $\mu$, which we will specify later. The main difference between our approach and the approach
of Ref. \onlinecite{Yue1993} is that we find an asymptotically exact solution of Eq. (\ref{interaction}),
while in Ref. \onlinecite{Yue1993} it was solved perturbatively.  The reason why asymptotically exact solution can be found is that the amplitude of $V_H(z)$ falls off slowly with $z$, so that the relevant
values of $q_{\s F}z$ are big. This, in turn, suggests
to search for $\psi_1(z)$ in the form

\begin{equation}
\label{psiH1}
\psi_1(z)=A_{+}(z)e^{iq_{\s F}z}+A_{-}(z)e^{-iq_{\s F}z},
\end{equation}
where the functions $A_{+}$ and $A_{-}$ change slowly with $z$, so that their second derivatives can be neglected.
Upon substituting Eq. (\ref{psiH1}) into Eq. (\ref{interaction}), and neglecting non-resonant terms $\exp(\pm 3iq_{\s F}z)$,
we arrive to a coupled system of the first-order equations
\begin{eqnarray}
\label{firstorder1}
-2iq_{\s F}\frac{\partial A_{+}(z)}{\partial z}+\frac{\mu}{2z}A_{-}(z)=\left(E+1-q_{\s F}^2\right)A_{+}(z),\nonumber\\
2iq_{\s F}\frac{\partial A_{-}(z)}{\partial z}+\frac{\mu}{2z}A_{+}(z)=\left(E+1-q_{\s F}^2\right)A_{-}(z).\hspace{2mm}
\end{eqnarray}
It appears that this system can be solved {\em exactly} for arbitrary interaction strength, $\mu$.
To see this, we first perform a rescaling
\begin{equation}
\label{rescaling}
y=z\left(\frac{E+1-q_{\s F}^2}{2q_{\s F}}\right),
\end{equation}
and then introduce the auxiliary functions
\begin{equation}
\label{a+ib}
a(y)=A_{+}(y)+iA_{-}(y),~b(y)=A_{+}(y)-iA_{-}(y).
\end{equation}
Then the system Eq. (\ref{firstorder1}) reduces to
\begin{eqnarray}
\label{AB}
\frac{\partial a}{\partial y}+\frac{\mu}{4q_{\s F}y}a(y)=i b(y),\nonumber\\
\frac{\partial b}{\partial y}-\frac{\mu}{4q_{\s F}y}b(y)=ia(y).
\end{eqnarray}
In the rescaled form, the system contains a single dimensionless parameter, $\frac{\mu}{4q_{\s F}}$. As a next step, we substitute $b(y)$ from the first equation into the second equation and arrive to the following second-order differential equation
\begin{equation}
\label{bessel}
\frac{\partial^2 a}{\partial y^2}+\Bigg[1+\frac{1-4(\frac{\mu}{4q_{\s F}}+\frac{1}{2})^2}{4y^2} \Bigg]a(y)=0.
\end{equation}
The general solution of this equation can be presented as a linear combination
\begin{equation}
\label{besselsolution}
a(y)=y^{1/2}\Big[c_{\s 1}J_{\s \frac{\mu}{4q_{\s F}}+\frac{1}{2}}(y)+c_{\s 2}J_{\s -\frac{\mu}{4q_{\s F}}-\frac{1}{2}}(y)\Big],
\end{equation}
where $J_{\s \frac{\mu}{4q_{\s F}}+\frac{1}{2}}$ and $J_{\s -\frac{\mu}{4q_{\s F}}-\frac{1}{2}}$ are the Bessel functions.
At large $y$ both Bessel functions oscillate, so that
the value of the transmission coefficient is governed by the ratio $c_{\s 1}/c_{\s 2}$. This ratio is determined by the condition that at small $y=y_c$, where the Friedel oscillations are terminated (see Appendix A), the amplitude of the reflected wave vanishes. The final expression for the transmission coefficient, reads
\begin{multline}
\label{transmissionbessel1}
t_{\textit{\tiny{Bragg}}}=\frac{\left(2\pi y_c\right)^{1/2}J_{\s \frac{\mu}{4q_{\s F}}-\frac{1}{2}}(y_c)J_{\s -\frac{\mu}{4q_{\s F}}-\frac{1}{2}}(y_c)}{J_{\s \frac{\mu}{4q_{\s F}}-\frac{1}{2}}(y_c)e^{i\frac{\pi\mu}{8q_{\s F}}}+J_{\s -\frac{\mu}{4q_{\s F}}-\frac{1}{2}}(y_c)e^{-i\frac{\pi\mu}{8q_{\s F}}}}.
\end{multline}
The details of the derivation are presented in the Appendix B.

The result Eq. (\ref{transmissionbessel1}) can be simplified when $y_c$ is small.
Then we can use the small-argument asymptotes of the Bessel functions and obtain
\begin{equation}
\label{cosh}
t_{\textit{\tiny{Bragg}}}=\frac{1}{\cosh\left(\frac{\mu}{4q_{\s F}}\ln y_c \right)}.
\end{equation}
In deriving this expression we took into account that the interactions are
weak in the usual sense, namely that the typical interaction energy is much
smaller than the Fermi energy. This condition ensures that $\frac{\mu}{q_{\s F}}$ is small.

Concerning the value of $y_c$, in Appendix A it is demonstrated that the Friedel
oscillations are terminated at $z=z_c\sim \frac{q_{\s 0}}{\Gamma}$. Using the relation Eq. (\ref{rescaling}), we find that, within a numerical factor, $y_c$ is given by
\begin{equation}
\label{yc}
y_c=\frac{E-E_{\s F}}{\Gamma}.
\end{equation}
We see that in the interesting limit when the Fermi level is close to the resonance
$y_c$ is indeed small.

Equations (\ref{cosh}) and (\ref{yc}) describe how the transmission of the
Bragg mirror evolves with energy.
Indeed, the argument of the hyperbolic cosine is the product of a small factor $\frac{\mu}{4q_{\s F}}$ and a big factor $\ln y_c$. If this product is small, e.g. when the interactions are weak, then the transmission coefficient is close to $1$. On the contrary, if
the product is big, we have
\begin{equation}
\label{tFreflective}
t_{\textit{\tiny{Bragg}}}=\left(\frac{2|E-E_{\s F}|}{\Gamma}\right)^{\frac{|\mu|}{4q_{\s F}}} \ll 1,
\end{equation}
i.e. the mirror is highly reflective.

 To conclude this Section, we present the microscopic expression for the parameter $\mu$ in terms of the Fourier components of the interaction potential. This expression follows from the expression for the amplitude of the oscillations of the electron density, calculated in Appendix A, and has the form
\begin{equation}
\label{mudefine}
\mu=\frac{\nu q_{\s F}}{2}|r_1(E_{\s F})|^2,
\end{equation}
where $\nu$ is given by
\begin{equation}
\label{nu}
\nu=\frac{V(0)-V(2q_{\s F})}{2\pi \hbar v_{\s F}}.
\end{equation}
The term $V(0)$ comes from the exchange potential, while $V(2q_{\s F})$
comes from the Hartree potential; $v_{\s F}$ stands for the Fermi velocity.

Note that the transmission, $t_{\textit{\tiny{Bragg}}}$, is full not only
in the absence of electron-electron interactions. If the interactions are present,
but there is no reflection from the impurity, $r_1(E_{\s F})=0$,
then transmission is also full. This is natural, since in the absence of reflection,
the Friedel oscillations do not form.

\section{Energy dependence of the effective reflection}
In Eq. (\ref{teffavg}) both $t_1$ and $t_{\textit{\tiny{Bragg}}}$ are the functions
of energy. While $t_1$ is a growing function of energy, $t_{\textit{\tiny{Bragg}}}$ grows
with increasing $|E-E_{\s F}|$. In addition, the power, $\frac{\mu}{4q_{\s F}}$,
in Eq. (\ref{tFreflective}) depends on the difference $|E_{\s 0}-E_{\s F}|$, see Appendix~A.

\begin{figure}
\includegraphics[scale=0.32]{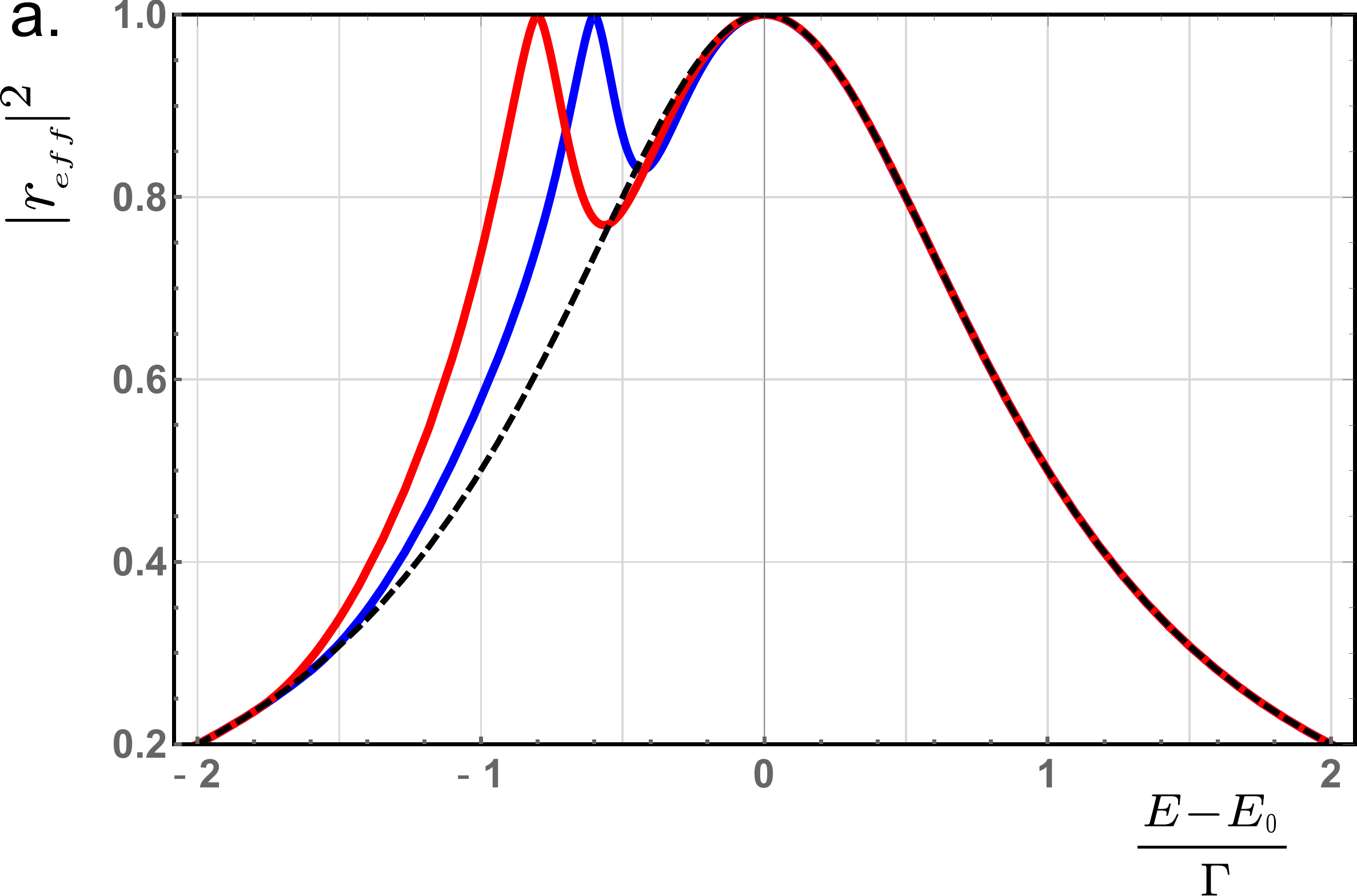}
\includegraphics[scale=0.32]{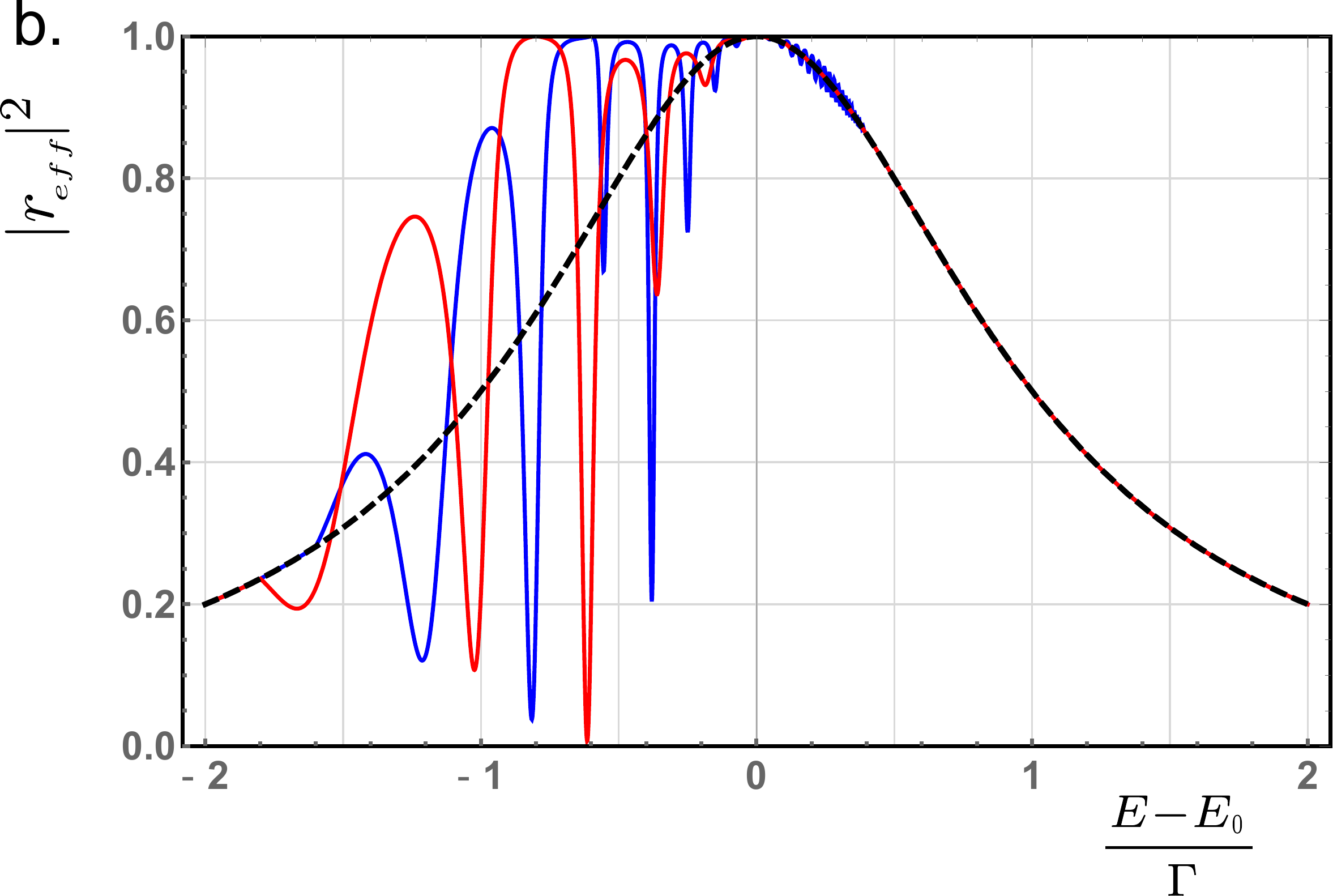}
\caption{(Color online) (a) In the absence of interactions, the effective power reflection coefficient is a Lorentzian, $|r_{\s eff}|^2=\left[1+\frac{(E-E_{\s 0})^2}{\Gamma^2}\right]^{-1}$
 (black dashed line).
 With interactions, full reflection takes place at two energies
 $E=E_{\s 0}$ as a result of scattering from the
 impurity and at $E=E_{\s F}$  as a result of scattering from the Bragg mirror. This is illustrated by red and blue
curves plotted from Eqs. (\ref{reffavg}) and (\ref{cosh})
for $(E_{\s 0}-E_{\s F})=0.8\Gamma$ and $(E_{\s 0}-E_{\s F})=0.6\Gamma$, respectively. The interaction strength in
both curves is chosen to be $\frac{\mu}{4q_{\s F}}=0.4$. (b) Scattering by two Bragg mirrors can, for certain energies, transform the resonant reflection into the resonant transmission. While the plot (a) shows the average over the phase, $\beta$, the plot (b) shows the reflection profile for the same parameters {\em prior to averaging}.  
  }
\label{Energydependence}
\end{figure}

Concerning the overall dependence $|r_{\s eff}(E)|^2$, the situation is most transparent when the Fermi level lies away from the resonance.
Then the presence of the Bragg mirrors manifests itself only near $E=E{\s F}$. Bragg
mirrors cause a spike in the reflection.
When the spacing between $E_{\s F}$ and $E_{\s 0}$ is much smaller than the width
of the resonance, there are two features in  $|r_{\s eff}(E)|^2$-dependence that are
present for any interaction strength.
Firstly, the reflection is full for any position of the Fermi
level when the energy of the incident electron is $E=E_{\s 0}$. This is because the electron is fully
reflected even in the absence of the Friedel oscillations.
Secondly, $|r_{\s eff}(E)|^2=1$ at $E=E_{\s F}$ due to full reflection from the mirror. Thus, in the domain $-E_{\s F}<E<0$, the reflection coefficient should pass through a minimum.  Indeed, this minimum is present in
the curves $|r_{\s eff}(E)|^2$ plotted from Eqs. (\ref{reffavg}) and (\ref{cosh})
in Fig. \ref{Energydependence}.

\section{Discussion}
({\em i}) To establish the relation between our results and those
obtained within the renormalization-group
approach\cite{Yue1993,Yue1994,Maslov,Nagaev,Nazarov2003,Gornyi2003}
we assume that the reflection of the Bragg mirrors is weak and expand Eq.  (\ref{reffabs})
with respect to $|r_{\textit{\tiny{Bragg}}}|^2$. This yields
\begin{equation}
\label{expanded}
|r_{\s eff}|^2-|r_1|^2=4\left( 1-|r_1|^2\right)\Big[|r_{\textit{\tiny{Bragg}}}|^2+|r_1||r_{\textit{\tiny{Bragg}}}|\cos\beta\Big].
\end{equation}
The second term in the brackets contains the first power of $|r_{\textit{\tiny{Bragg}}}|$, unlike the first term
 which contains $|r_{\textit{\tiny{Bragg}}}|^2$. This second term comes from interference of incident and reflected waves
 passing through the Bragg mirror. If we average Eq. (\ref{expanded}) over $\beta$, the second term
 will disappear. Then it is the first term, $1-t_{\textit{\tiny{Bragg}}}^2$, that will describe the reduction of the transmission of the impurity
due to electron-electron interactions. As follows from Eq. \ref{cosh},
$|r_{\textit{\tiny{Bragg}}}|^2$ is proportional to $|r_1|^2$
and contains $\mu\ln y_c$. Then Eq. (\ref{expanded})
 reproduces the main result of Ref. \onlinecite{Yue1993}. In Ref. \onlinecite{Yue1993} this result is subsequently converted to the renormalization-group equation.
We studied the limit in which both $|r_1|$ and $|r_{\textit{\tiny{Bragg}}}|$ are close to $1$. Then
the denominator in Eq. (\ref{reffabs}) is close to zero when $\cos \beta =-1$.
Definitely, the expansion with respect to $|r_{\textit{\tiny{Bragg}}}|$ and subsequent summation of the leading terms,
which is the essence of the renormalization-group approach, does not capture this resonant transmission.

({\em ii}) Adopting of the renormalization group approach in Refs. \onlinecite{Yue1993,Yue1994,Maslov,Nagaev,Nazarov2003,Gornyi2003}
relies on the assumption that the coefficients of the expansion of $|t_{\s eff}|^2$
is powers of $\ln(|E-E_{\s F}|)$ fall off as $\frac{1}{n!}$. Our calculation
is equivalent to the summation of all the orders of the expansion and confirms this assumption.

({\em iii}) The form Eq. (\ref{reflection}) of the resonant reflection is the same as
for the resonant tunneling between the two electrodes via a localized state located between
the electrons.  This suggests the interpretation of the resonant transmission as resonant tunneling
{\em between left-moving and right-moving electrons}. If this interpretation is correct,
the width, $\Gamma$, calculated from the golden rule should coincide with Eq. (\ref{Gamma}), and,
in particular, should be proportional to $U_0^3$. Taking into account that the normalized wave function
of the localized state has the form $\psi_2(z)=\kappa^{1/2}\exp(-\kappa|z|)$, the matrix element of
$\alpha\frac{\partial}{\partial z}$ between $\psi_2(z)$ and the right-moving
 plane wave, $\exp(iqz)$,   is given by
\begin{equation}
\label{overlap}
i\alpha\kappa^{1/2}q\int\limits_{-\infty}^{\infty}dz\exp\Big[iqz-\kappa|z|\Big]=2i\frac{q\kappa^{3/2}}{q^2+\kappa^2}.
\end{equation}
One can neglect $\kappa^2$ in the denominator. Then the square of the matrix element is proportional to
$\kappa^3$ and thus to $U_0^3$, since, at resonance, $\kappa=\frac{U_0}{2}$.

({\em iv}) There is a question whether the attenuation of electron wave functions upon passage
of the Bragg mirrors disturbs the shape of the Friedel oscillations. It is important that this
disturbance is negligible. Qualitatively, this follows from the fact that many states with $E<E_{\s F}$
are responsible for the formation of the Bragg mirrors, while only the states with
$|E-E_{\s F}| \lesssim \nu\Gamma$ are strongly affected by the Bragg mirrors.

({\em v}) Another question is why we did not take into account the Friedel oscillations originating from the electron reflection within the same sub-band. Indeed, while the Friedel oscillations caused by
the resonant reflection develop at large distances $z_c \sim \frac{q_0}{\Gamma}$, ``non-resonant" Friedel oscillations start at much smaller $z\sim 1$. To answer this question one should estimate the
contribution to the reflection coefficient within the domain $1<z<z_c$, where non-resonant Friedel oscillations dominate.
The amplitude of the these oscillations is $\sim \frac{U_0}{q_{\s F}}$ and they fall off as $1/z$. This leads to the estimate $\frac{U_0}{q_{\s F}}\ln(z_c)$ as in Ref. \onlinecite{Yue1993}. Since
$\frac{U_0}{q_{\s F}}\ll 1$, the weakness of non-resonant reflection cannot be compensated by the logarithmically big factor $\ln \left(\frac{q_0}{\Gamma}\right)$, it is for this reason we have neglected the Friedel oscillations originating from the reflection within the same sub-band.

({\em vi}) Our main finding is that, for weak transmission through a single Bragg mirror, the
net transmission from two Bragg mirrors and the impurity can be close to one. This
enhancement of the net transmission takes place when the ``Fabry-Perot" condition $\cos\beta \approx -1$ is met. Then the denominator in Eq.  (\ref{reffabs}) becomes small.  This happens near certain distinct
energies of incident electron. Averaging over the phase, $\beta$, employed above, requires that  there are many such energies within the interval $|E_0-E_{\s F}|$. To verify that this is the case, consider
the contribution to $\beta$ coming from the factor $\exp(iq_{\s F}z)$ in Eq. (\ref{psiH1}).
As an estimate for $z$ in this factor one should take the effective length of the Bragg mirror
where the reflection is formed. From Eq. (\ref{bessel}) we see that this length is determined by the condition $y\gg 1$. At these values of $y$ the product
 $y^{1/2}J_{\s \frac{\mu}{4q_{\s F}}+\frac{1}{2}}(y)$ saturates meaning that the formation of the
  Bragg reflection is complete.  The condition $y\gg 1$ transforms into
  the condition $z \gtrsim \frac{q_{\s F}}{E-E_{\s F}}$.
Thus, the contribution
to $\beta$ from the phase, $ \Phi_{\textit{\tiny{Bragg}}}   $, accumulation in the course of traveling through the mirror is of the order
of $(E-E_{\s F})^{-1}$. In the relevant domain $|E_0-E_{\s F}|\lesssim \Gamma$ this phase goes through
$(2n+1)\pi$ many times.

\vspace{4mm}

\centerline{\bf Acknowledgements}

We are strongly grateful to E. G. Mishchenko for a number
of illuminating discussions.
The work was supported by the Department of
Energy, Office of Basic Energy Sciences, Grant No. DE-
FG02-06ER46313.
\appendix
\section{Magnitude of the Friedel Oscillations}
The scattering of electrons from the impurity modifies the electron densities around the impurity. In the presence of electron-electron interaction this modulation of density leads to an additional scattering, which we call ``Bragg mirror" in the main text. This scattering barrier is also called Hartree potential,
\begin{equation}
\label{Hartree}
V_H(z)=\int\limits_{-\infty}^{\infty} V(z-y) \delta n(y)dy,
\end{equation}
where $V(z-y)$ is the interaction potential and $\delta n(y)$ is the fluctuation of the density. Assuming interaction to be short ranged, $V(z-y)=\nu~\delta(z-y)$, we see that the Hartree potential takes the form, $V_H(z)=\nu~\delta n(z)$. Now, the modulation of the electron density, $\delta n(z)$, which depends on the reflection coefficient, $r_1$, reads
\begin{multline}
\label{modulation}
\delta n(z)=\int\limits_{0}^{q_{\s F}} \frac{dq}{\pi} ~2 \text{Re}(r_1(q)~e^{2iqz})\\=\hspace{-2mm}
\int\limits_{0}^{q_{\s F}} \frac{dq}{\pi} \frac{\Gamma}{\left[\Gamma^2 +\left(q_{\s 0}^2-q^2\right)^2\right]^{1/2}} \cos\left(2q|z|+\tan^{-1}\hspace{-1mm}\frac{\Gamma}{q_{\s 0}^2-q^2}\right),
\end{multline}
where $q_{\s 0}=(1+E_{\s 0})^{1/2}$, see Eq. (\ref{q}). Upon measuring $q$ from $q_{\s F}$ and introducing new variables,
\begin{equation}
\label{u}
u=2q_{\s 0}\frac{q_{\s F}-q}{\Gamma}, ~~ u_{\s 0}=2q_{\s 0}\frac{q_{\s 0}-q_{\s F}}{\Gamma},\hspace{-3mm}
\end{equation}
Eq. (\ref{modulation}) assumes the form
\begin{multline}
\label{modulation1}
\delta n(z)=\frac{\Gamma}{2\pi q_{\s 0}}\int\limits_{0}^{\frac{2q_{\s\s 0}q_{\s F}}{\Gamma}} du~\frac{1}{\Big[1+\left(u+u_{\s 0}\right)^2\Big]^{1/2}} \\
\times\cos\Bigg[2|z|\left(q_{\s F}-\frac{\Gamma}{2q_{\s 0}}u +\tan^{-1} \frac{1}{u+u_{\s 0}}\right) \Bigg].
\end{multline}
It is convenient to separate the contributions proportional to $\sin(2q|z|)$ and to $\cos(2q|z|)$. This yields
\begin{multline}
\label{modulation3}
\delta n(z)=\frac{\Gamma}{2\pi q_{\s 0}}\int\limits_{0}^{\frac{2q_{\s\s 0}q_{\s F}}{\Gamma}} du~\frac{1}{1+\left(u+u_{\s 0}\right)^2}\times
\\
 \Bigg\{\hspace{-1mm}\left(u+u_{\s 0}\right)
\cos\!\left[2|z|\left(q_{\s F}-\frac{\Gamma}{2q_{\s 0}}u\right)\right]
-\sin\! \left[2|z|\left(q_{\s F}-\frac{\Gamma}{2q_{\s 0}}u\right)\right]\hspace{-1.5mm}\Bigg\}.\hspace{-2mm}
\end{multline}
The shift, $\frac{\Gamma}{2q_{\s 0}}u$, of the arguments of both cosine and sine leads to the factors $\sin\left(\frac{\Gamma|z|}{q_{\s 0}}u\right)$ and $\cos\left(\frac{\Gamma|z|}{q_{\s 0}}u\right)$ in the numerator.
For $\frac{\Gamma|z|}{q_{\s 0}}\gg 1$, both terms rapidly oscillate with $u$. Without $u$-dependence of the prefactor, the contribution from the cosine term will vanish. With the prefactor the contribution of this term remains much smaller than the contribution of the sine term. Retaining only the sine-term we get
\begin{equation}
\label{leading}
\delta n(z)=\cos\left(2q_{\s F} |z|\right)\frac{\Gamma}{2\pi q_{\s 0}}\int\limits_{0}^{\frac{2q_{\s 0}q_{\s F}}{\Gamma}} du~\frac{\sin\left(\frac{\Gamma|z|}{q_{\s 0}}u\right)}{1+\left(u+u_{\s 0}\right)^2 }.
\end{equation}
For $\frac{\Gamma|z|}{q_{\s 0}}\gg 1$ we can replace the upper limit of the integral by infinity and neglect the $u$-dependence of the denominator. This leads to the final answer
\begin{equation}
\label{limit2}
\delta n(z)=\frac{|r_1(E_{\s F})|^2}{2\pi|z|}\cos\left(2q_{\s F} |z|\right),
\end{equation}
where we have used the fact that $|r_1(E_{\s F})|^2$ is $(1+u_{\s 0}^2)^{-1}$.
Note that, unlike the conventional Friedel oscillations\cite{Yue1993}, Eq. (\ref{limit2}) contains the second power of $|r_1(E_{\s F}|$. Extra power originates
from the phase of the cosine in Eq. (\ref{modulation1}), which is strongly energy-dependent.

The most important outcome of the above analysis is that the Friedel oscillations are terminated at rather large distances $z=z_c\sim\frac{q_{\s 0}}{\Gamma}$. We have used this value as a cutoff of log-divergence in the main text.


\section{Calculation of transmission coefficient from more rigorous approach}
Substituting the general form  Eq. (\ref{besselsolution}) of $a(y)$ in the
system Eq. (\ref{AB}) we find the following general form of $b(y)$  \begin{equation}
\label{besselsolutionb}
b(y)=-iy^{1/2}\Big[c_{\s 1}J_{\s \frac{\mu}{4q_{\s F}}-\frac{1}{2}}(y)-c_{\s 2}J_{\s -\frac{\mu}{4q_{\s F}}+\frac{1}{2}}(y)\Big].
\end{equation}
Once $a(y)$ and $b(y)$ are known, the incident amplitude,
$A_{+}(y)=\frac{1}{2}[a(y)+b(y)]$, and the reflected amplitude
$A_{-}(y)=\frac{1}{2i}[a(y)-b(y)]$ can be expressed as a combination of the Bessel functions
\begin{eqnarray}
\label{A+A-}
A_{+}=\frac{y^{1/2}}{2}\Bigg\{c_{\s 1}\Big[J_{\s \frac{\mu}{4q_{\s F}}+\frac{1}{2}}(y)-iJ_{\s \frac{\mu}{4q_{\s F}}-\frac{1}{2}}(y)\Big]\nonumber\\
+c_{\s 2}\Big[J_{\s -\frac{\mu}{4q_{\s F}}-\frac{1}{2}}(y)+iJ_{\s -\frac{\mu}{4q_{\s F}}+\frac{1}{2}}(y)\Big]\Bigg\},\\
A_{-}=\frac{y^{1/2}}{2i}\Bigg\{c_{\s 1}\Big[J_{\s \frac{\mu}{4q_{\s F}}+\frac{1}{2}}(y)+iJ_{\s \frac{\mu}{4q_{\s F}}-\frac{1}{2}}(y)\Big]\nonumber\\
+c_{\s 2}\Big[J_{\s -\frac{\mu}{4q_{\s F}}-\frac{1}{2}}(y)-iJ_{\s -\frac{\mu}{4q_{\s F}}+\frac{1}{2}}(y)\Big]\Bigg\}.
\end{eqnarray}
In the limit $y\rightarrow \infty$, the behavior of
 $A_{+}$ and $A_{-}$ is the following
\begin{eqnarray}
\label{largeasymptotes}
A_{+}=\frac{1}{(2\pi)^{1/2}}\Big[c_{\s 2}e^{i\frac{\pi\mu}{8q_{\s F}}}-ic_{\s 1}e^{-i\frac{\pi\mu}{8q_{\s F}}}\Big]e^{iy},\nonumber\\
A_{-}=\frac{-i}{(2\pi)^{1/2}}\Big[c_{\s 2}e^{-i\frac{\pi\mu}{8q_{\s F}}}+ic_{\s 1}e^{i\frac{\pi\mu}{8q_{\s F}}}\Big]e^{-iy}.
\end{eqnarray}
For small $y$, we have $J_{\s \pm\frac{\mu}{4q_{\s F}}+\frac{1}{2}}(y) \ll J_{\s \pm\frac{\mu}{4q_{\s F}}-\frac{1}{2}}(y)$, so the asymptotic expressions for $A_{+}$ and $A_{-}$ can be written as
\begin{eqnarray}
\label{smallasymptotes}
A_{-}=\frac{y^{1/2}}{2i}\Big[ic_{\s 1}J_{\s \frac{\mu}{4q_{\s F}}-\frac{1}{2}}(y)+c_{\s 2}J_{\s -\frac{\mu}{4q_{\s F}}-\frac{1}{2}}(y)\Big],\nonumber\\
A_{+}=\frac{y^{1/2}}{2}\Big[-ic_{\s 1}J_{\s \frac{\mu}{4q_{\s F}}-\frac{1}{2}}(y)+c_{\s 2}J_{\s -\frac{\mu}{4q_{\s F}}-\frac{1}{2}}(y)\Big]
\end{eqnarray}
To find the transmission of the Bragg mirror we need to know
the ratio $c_{\s 1}/c_{\s 2}$. This ratio is determined by the
condition that the Bragg mirror exists only for $y>y_c$.
Correspondingly, the amplitude $A_{-}$ at $y=y_c$ is zero.
This yields
\begin{equation}
\label{c1c2ratio}
\frac{c_{\s 1}}{c_{\s 2}}=i\frac{J_{\s -\frac{\mu}{4q_{\s F}}-\frac{1}{2}}(y_c)}{J_{\s \frac{\mu}{4q_{\s F}}-\frac{1}{2}}(y_c)}.
\end{equation}
By definition, the amplitude transmission coefficient of the mirror, $t_{\textit{\tiny{Bragg}}}$,
is the ratio of the values of $A_{+}$ at $y=y_c$ and at large $y$.
Using the ratio Eq. (\ref{c1c2ratio}) and Eqs. (\ref{largeasymptotes}), (\ref{smallasymptotes})  we arrive to Eq. (\ref{transmissionbessel1}) of the main text.

\section{Alternative derivation of resonant reflection}
It is instructive to trace how the resonant reflection of
$\uparrow$ electrons emerges from the closed equation for
the spin component $\psi_1(z)$. To derive this equation, we introduce the Fourier transform,
\begin{equation}
\label{Fourier1}
\varphi_2(p)=\frac{1}{2\pi}\int\limits_{-\infty}^{\infty}dz~\psi_2(z)\exp(-ipz),
\end{equation}
we rewrite the second equation of the system Eq. (\ref{System2}) in the form
\begin{equation}
\label{Fourier2}
(p^2+\kappa^2)\varphi_2(p)+\frac{U_0}{2\pi}\psi_2(0)=-\frac{\alpha}{2\pi}\int\limits_{-\infty}^{\infty}dz~
\frac{\partial \psi_1}{\partial dz}\exp(-ipz).
\end{equation}
Expressing $\varphi_2(p)$ and substituting it into the self-consistency condition
\begin{equation}
\label{Fourier3}
\psi_2(0)=\int\limits_{-\infty}^{\infty}dp~\varphi_2(p),
\end{equation}
we find
\begin{equation}
\label{Fourier4}
\psi_2(0)=-\frac{\alpha}{U_0+2\kappa}\int\limits_{-\infty}^{\infty}dz~\frac{\partial \psi_1}{\partial z}e^{-\kappa|z|}.
\end{equation}
Substituting Eq. (\ref{Fourier4}) into Eq. (\ref{Fourier2}),
we express $\varphi_2(p)$ in terms of $\psi_1(z)$
\begin{equation}
\label{Fourier5}
\varphi_2(p)=-\frac{\alpha}{2\pi(p^2+\kappa^2)}
\Biggl[\int\limits_{-\infty}^{\infty}dz\frac{\partial \psi_1}{\partial z}\Bigl(e^{-ipz}-
\frac{U_0}{U_0+2\kappa} e^{-\kappa|z|}\Bigr)\Biggr].
\end{equation}
Multiplying Eq. (\ref{Fourier5}) by $\exp(ipz)$ and integrating over $p$,
we get the following expression for $\psi_2(z)$
\begin{equation}
\label{Fourier6}
\psi_2(z)=\frac{\alpha}{2\kappa}\Biggl[-\!\!\int\limits_{-\infty}^{\infty}\!
\!dz_1\frac{\partial \psi_1}{\partial z_1}
e^{-\kappa|z-z_1|}+\frac{U_0e^{-\kappa|z|}}{U_0+2\kappa}\int\limits_{-\infty}^{\infty}\!\!dz_1\frac{\partial \psi_1}{\partial z_1}
e^{-\kappa|z_1|}    \Biggr].
\end{equation}

\begin{widetext}
\begin{equation}
\label{selfconst1}
-\frac{\partial^2\psi_1}{\partial z^2}+U_0 \delta(z)\psi_1 - (E +1)\psi_1=\frac{\alpha^2}{2\kappa}\frac{\partial}{\partial z}\left[\int\limits_{-\infty}^{\infty} dz_1  \frac{\partial \psi_1}{\partial z_1}e^{-\kappa |z-z_1|} -\frac{U_0e^{-\kappa |z|}}{U_0 +2\kappa} \left(\int\limits_{-\infty}^{\infty} dz_1  \frac{\partial \psi_1}{\partial z_1}e^{-\kappa |z_1|} \right)\right].
\end{equation}
\end{widetext}
The term responsible for the resonant reflection is the second term in the right-hand
side. Near the resonance, it is much bigger than the first term. The term $U_0\delta(z)$ in the left-hand side describes a non-resonant scattering from the
impurity. Neglecting these terms we get
\begin{widetext}
\begin{equation}
\label{selfconst2}
-\frac{\partial^2\psi_1}{\partial z^2} - (E+1)\psi_1=\frac{\alpha^2}{2}\frac{U_0}{U_0 +2\kappa} \left(\int\limits_{-\infty}^{\infty} dz_1  \frac{\partial \psi_1}{\partial z_1}e^{-\kappa |z_1|} \right)e^{-\kappa |z|}\sign(z).
\end{equation}
\end{widetext}
We see that the right-hand side is a {\em discontinuous} function
of $z$. This fact constitutes the origin of the resonant reflection.
For example, if we integrate Eq. (\ref{selfconst2}) near $z=0$, we will
see that, unlike conventional scattering, the derivative, $\frac{\partial\psi_1}{\partial z}$, is continuous at the position of impurity.
This translates into the relation $t_1=1-r_1$, which is nothing but
Eq. (\ref{t1r1}). To derive the second equation, Eq. (\ref{t1-r1}),
one should notice that $\psi_1(z)$ is present in the right-hand side only
 under the integral, so that the explicit solution of Eq. (\ref{selfconst2})
 can be readily found. This solution also contains $t_1$ and $r_1$. Then
Eq. (\ref{t1-r1}) emerges as a self-consistency condition.

\section{Smallness of the transmission through the Bragg mirror}
The fact that the transmission coefficient, $t_{\textit{\tiny Bragg}}$, is small suggests to
use the semiclassical approach to calculate $t_{\textit{\tiny Bragg}}$.
Semiclassical approach is equivalent to the assumption that $A_{+}$ and $A_{-}$, which are the solutions of the system Eq. (\ref{firstorder1}) are proportional to $\exp\left[\pm S(z)\right]$,
where $S(z)$ is the action. From the system Eq. (\ref{firstorder1}) we find
\begin{equation}
\label{action}
\frac{dS}{dz}=\frac{1}{2q_{\s F}}\Bigg[\frac{\mu^2}{4z^2}-\left(E+1-q_{\s F}^2 \right)^2 \Bigg]^{1/2}.
\end{equation}
It is seen from Eq. (\ref{action}) that the functions $A_{\pm}$ oscillate at $z>z_t$, where the turning point $z_t$ is given
by
\begin{equation}
\label{turningpoint}
z_t=\frac{|\mu|}{2 |E+1-q_{\s F}^2 |}.
\end{equation}
For smaller $z$, $A_{\pm}(z)$ are the combinations of growing and decaying exponents. This behavior is sustained
in the interval $z_c<z<z_t$, where $z_c\sim 1/{\Gamma}$ is the point where the Friedel oscillations are terminated (see Appendix A).
For applicability of the semiclassics, the action
\begin{equation}
\label{action1}
S(z_t)-S(z_c)=\frac{1}{2q_{\s F}}\int\limits_{z_c}^{z_t}dz~\Bigg[\frac{\mu^2}{4z^2}-|E+1-q_{\s F}^2 |^2 \Bigg]^{1/2}
\end{equation}
accumulated between the points $z_c$ and $z_t$ should be much bigger than one. However, the evaluation of the integral suggests that this condition reduces to $|\mu|/4q_{\s F}\ln(z_t/z_c)\gg 1$, which is not the case for weak electron-electron interactions. This is why we derived $t_{\textit{\tiny Bragg}}$ from the exact solution of the system Eq. (\ref{firstorder1}). Failure of the semiclassics
can be traced back to neglecting  the $z$-dependence of the prefactors $A_{+}$ and $A_{-}$.


\end{document}